\documentclass[12pt,preprint]{aastex}
\begin{document}
\title{Void-Supercluster Alignments}
\author{Daeseong Park and Jounghun Lee}
\affil{Department of Physics and Astronomy, FPRD, Seoul National University, 
Seoul 151-747, Korea} 
\email{pds2001@astro.snu.ac.kr, jounghun@astro.snu.ac.kr}
\begin{abstract}
We investigate alignments between the spin axes of cosmic voids and the 
principal axes of nearby superclusters using the Millennium Run simulation 
of a $\Lambda$CDM cosmology. The concept of void spin was first introduced 
by Lee and Park in 2006 to quantify the tidal effect on voids from the 
surrounding matter distribution. Our numerical analysis reveals that the 
void spin axes are strongly aligned with the supercluster minor axes, but 
anti-aligned with the major axes, and have no correlations with the 
intermediate axes. We provide physical explanations to this numerical results 
on the basis of tidally induced correlations.  It is expected that our 
work will provide a new insight into the characterization of the cosmic web 
on the largest scale.
\end{abstract}


\section{INTRODUCTION}

Voids stand for those regions in the universe which have extremely low 
density. There is still no consensus on how to define voids. In recent 
numerical approaches using N-body simulations, voids are often defined as 
regions empty of massive dark matter halos. In real observations, 
however, voids are usually defined as underdense regions with the mean density 
contrast close to $-0.9$ \citep[e.g.,][]{hoy-vog04}, given the limitation that we cannot 
observe very faint galaxies.  Adopting the latter definition of voids, one can say 
that voids are not empty of massive dark halos but have some structures in them. 
An interesting question is how the void structures evolve with time. 

The standard picture based on the cold dark matter paradigm explains that 
the voids originated from the local minima of the initial density fluctuation 
and have become more and more underdense via gravitational rarefaction in the 
subsequent evolution \citep{ick84}. Recent high-resolution simulations, 
however, have revealed that the evolution of the void structures is more 
complicated than naively predicted \citep[e.g.,][]{got-etal03,col-etal05}: 
Not all voids remain underdense but some fraction of them actually collapse 
into bound halos. Although the collapse of voids may be described 
qualitatively as the occurrence of the clouds-in-voids in the frame of the 
standard excursion set theory \citep{she-van04}, it has yet to be well 
understood under what conditions and how frequently the collapse of voids 
takes place.

A first suspicion goes to the tidal forces from the surrounding matter 
distribution \citep{sah-etal94}. As noted by \citet{sha-etal04, sha-etal06}, 
the voids should be severely disturbed by the tidal influence due to the 
low density. Considering this possibility, a plausible scenario is that when 
the tidal effect on a void wins over the gravitational rarefaction, the void 
will become squeezed out to collapse into a bound halo. A remaining question 
is how to quantify the tidal effect on voids and predict its consequences.

Very recently, \citet[][hereafter, LP06]{lee-par06} introduced a new concept, 
{\it the void spin angular momentum}, which is defined as 
\begin{equation}
\label{eqn:voidspin}
{\bf J} = \frac{1}{M_{V}}\sum_{\alpha}^{N_{V}}m_{\alpha}{\bf r}_{\alpha}
\times{\bf v}_{\alpha},
\end{equation}
where $N_{v}$ is the total number of halos belonging to a void, 
$m_{\alpha}$, ${\bf r}_{\alpha}$, and ${\bf v}_{\alpha}$ are the mass, 
the position, and the velocity of the $\alpha$-th halo belonging to a void 
and $M_{V}\equiv \sum_{\alpha}^{N_{V}}m_{\alpha}$. Here the positions and 
the velocities of the void halos are measured relative to the center 
of mass of the void halos. 

As mentioned by LP06, although the quantity defined in equation 
(\ref{eqn:voidspin}) is not a real spin angular momentum since voids  
are not bound system, the concept of void spin turned out to be a 
very effective measure of the tidal effect from the surrounding matter.
To show that the void spin angular momentum is tidally induced, LP06 
measured the spatial correlations between the spin axes of neighboring 
voids analyzing numerical data from high-resolution simulation, and  
found that the numerical results are in excellent agreement with the 
analytic predictions based on tidally generated correlations.

If the void spin angular momentum is really induced by the tidal 
effect from the surrounding matter, then there should exist correlations 
not only between the spin axes of neighboring voids but also between the 
spin axes of voids and the neighboring matter distribution. 
This cross correlation, if found to exist, should be a more direct 
indication of the tidal effect on voids from the surrounding matter.

Our goal here is to measure the cross-correlations between the void-spin 
axes and the axes of the neighboring superclusters using numerical data 
from high-resolution simulations and to explain it physically.
The superclusters are considered as the counterparts of voids since 
they are the largest bound structures, comparable to voids in size. 

The outline of this paper is summarized as follows.  In \S 2, the numerical 
data are analyzed and the signals of void-supercluster alignments are 
detected.  In \S 3, analytic model for the void-supercluster alignments 
is derived and compared with the numerical results. 
In \S 4, the implications and caveats of our work are discussed, 
and a final conclusion is drawn.

\section{NUMERICAL RESULTS}

\subsection{Measurements of Void Spins}

We analyzed the halo catalog extracted from the Millennium Run simulation 
\citep{spr-etal05}, which is now available at 
http://www.mpa-garching.mpg.de/millennium. 
The simulation is of a flat $\Lambda$CDM universe for which the values of 
the key parameters are given as $(\Omega_{m},\sigma_{8},h,n_{s}) = 
(0.25,0.9,0.73,1)$ where $\Omega_{m}$, $\sigma_{8}$, $h$ and $n_{s}$ represent 
the matter density parameter, the rms density fluctuation on the top-hat scale 
of $8h^{-1}$Mpc, the dimensionless Hubble parameter and the slope of the 
primordial power spectrum. 

For our analysis, we exclude those halos from the Millennium Run catalog 
which have too low particle numbers, as those low-particle number halos 
are just poorly sampled, heavily affected by noise (V. Springel in private 
communication). Technically, we set the cut-off particle number at $50$, 
a minimum particle number for defining a halo density profile. 

In accordance with the void-finder algorithm proposed by 
\citet[][hereafter, HV02]{hoy-vog02} to the Millennium Run halo catalog, 
we take the following steps to identify voids: First, we classify the halos 
according to the wall/field criterion, given as $l=\bar{d}_{3}+3\sigma/2$, 
where $\bar{d}_{3}$ is the average distance to the third nearest neighbor 
and $\sigma$ is its standard deviation. For the Millennium Run halos, the 
criterion distance is found to be $l=2.78h^{-1}$Mpc. Note that this value is 
slightly higher than the value $l=2.44h^{-1}$Mpc used in our previous work 
\citep{lee-par06}. It is because in our previous work the voids were 
identified not from the halo catalog but from the galaxy catalog.  
Second, we place the wall halos on cubic grid cells each of which has a 
linear size of $l$, counting the number of halos in each cell. Third, 
increasing the radii of empty spheres from the center of all empty grid 
cells till each of them include three wall halos on the surface, we 
determine the largest possible empty spheres. 

Finally, we detect voids by categorizing the overlapping empty spheres whose 
radii are greater than the minimum size threshold. The minimum void size are 
set at $6h^{-1}$Mpc which was found through statistical significance test in 
our previous work \citep{lee-par06}.  A total $20291$ voids are identified 
in the z=0 catalog, among which only $6430$ voids are found to have more than 
$30$ halos. The mean density contrast ($\bar{\delta}_{v}$) and the mean 
effective Eulerian radius ($\bar{R}_{E}$) of these $6430$ 
voids are found to be $\bar{\delta}_{v}=-0.9$ and 
$\bar{R}_{E}=12.22h^{-1}$Mpc, respectively. 

Considering only these large voids with more than $30$ halos, we calculate 
the spin angular momentum by equation (\ref{eqn:voidspin}). It is worth 
mentioning here that the direction of the void spin angular momentum might 
depend on the cut-off particle number for the halo selection. In our analysis, 
we include only those halos in the Millennium Run halo catalog which contain 
more than $50$ particles. One may suspect that a different cut-off particle 
number might yield different directions of the void spins.  

\begin{figure}
  \begin{center}
   \plotone{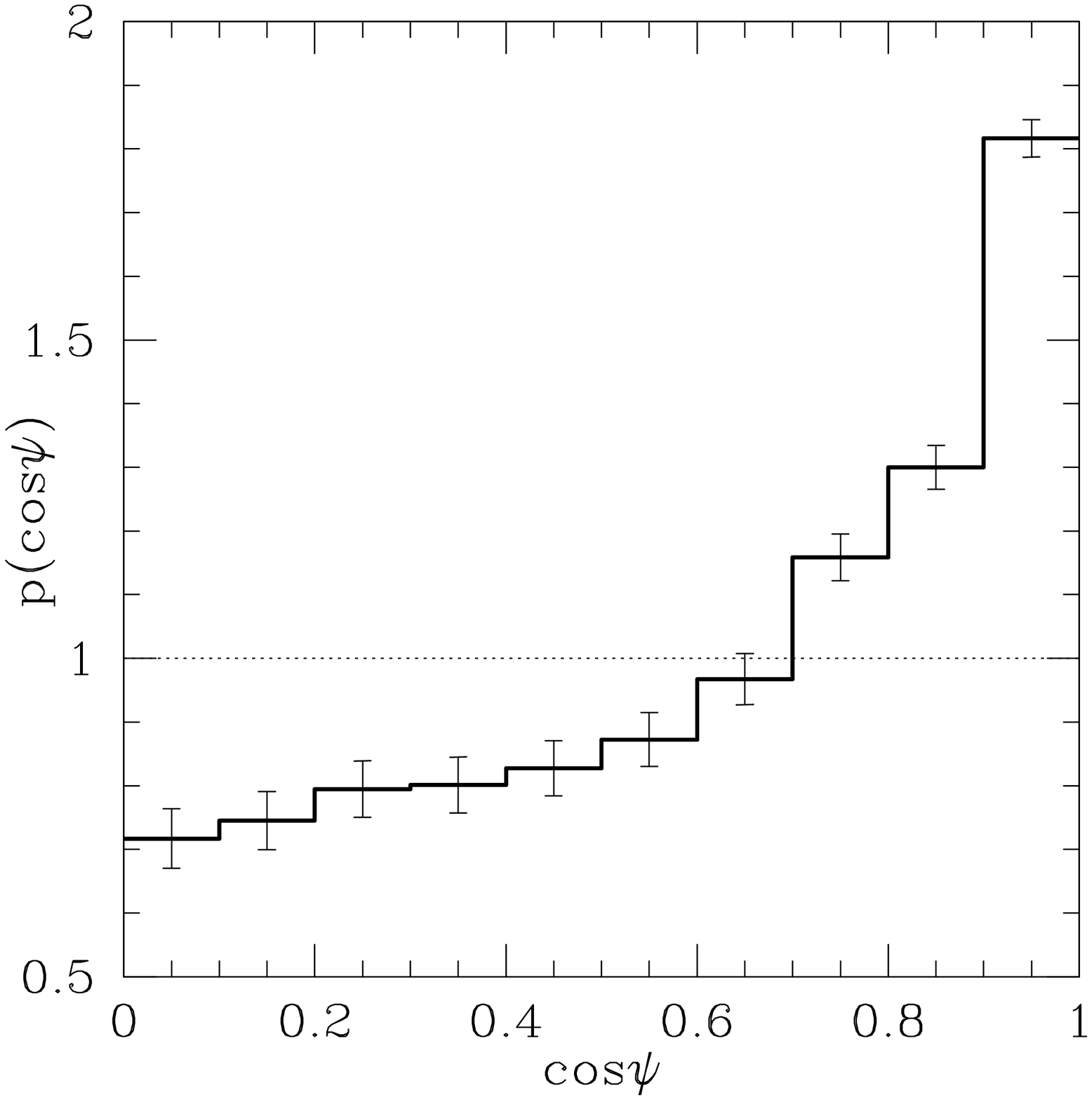}
\caption{The probability density distribution of the cosines of the angles 
between ${\bf J}$ and ${\bf J}^{\prime}$.}
\label{fig:psi}
 \end{center}
\end{figure}

To examine how the void spin angular momentum depends on the cut-off 
particle number, we calculate a second angular momentum ${\bf J}^{\prime}$ 
for each of the 6430 voids, taking all halos inside the void with more than 
$30$ particles, and calculate the cosines of the angles, $\psi$, between 
${\bf J}$ and ${\bf J}^{\prime}$ as
\begin{equation}
\cos\psi \equiv \frac{\vert {\bf J}\cdot{\bf J}^{\prime}\vert}
{\vert{\bf J}\vert\vert{\bf J}^{\prime}\vert}.
\end{equation}
Figure \ref{fig:psi} plots the probability distribution of $\cos\psi$ as 
histogram with the Poisson errors. The horizontal line corresponds to the 
case of no correlation. As can be seen, the two angular momentum vectors are 
very strongly aligned. It indicates that excluding those halos with lower 
particle number will not affect the void spin angular momentum significantly.

\subsection{Measurements of Supercluster Principal Axes}

A next task is to identify superclusters from the Millennium Run halo 
catalog. Following the common method \citep[e.g.,][]{wra-etal06}, 
we define a supercluster as a cluster of clusters, and identify superclusters 
with the help of the friends-of-friends algorithm (FOF). 
The cluster halos are selected as those halos whose mass exceeds a typical 
poor cluster mass, $M_{c}$. As in \citet{wra-etal06}, we set the value of 
$M_{c}$ at $1.75\times 10^{13}h^{-1}M_{\odot}$. 

Before applying the FOF algorithm to the Millennium Run halo catalog, one has 
to determine a linking length for the supercluster identification. To find an 
optimal linking length, $L$, we perform a statistical significance test by 
generate $100$ random Poisson samples of clusters which have the same number 
density in the same the box size as the Millennium halo catalog but having 
no clustering effect. The statistical significance of a supercluster can 
be estimated as $P(L) = 1 - N_{po}(L)/N_{sc}(L)$, where $N_{po}(L)$ and 
$N_{sc}(L)$ are the numbers of superclusters found in the random samples 
and in the Millennium Run simulation data at a linking length $L$, 
respectively.\citep{bas-etal06}.

\begin{figure}
  \begin{center}
   \plotone{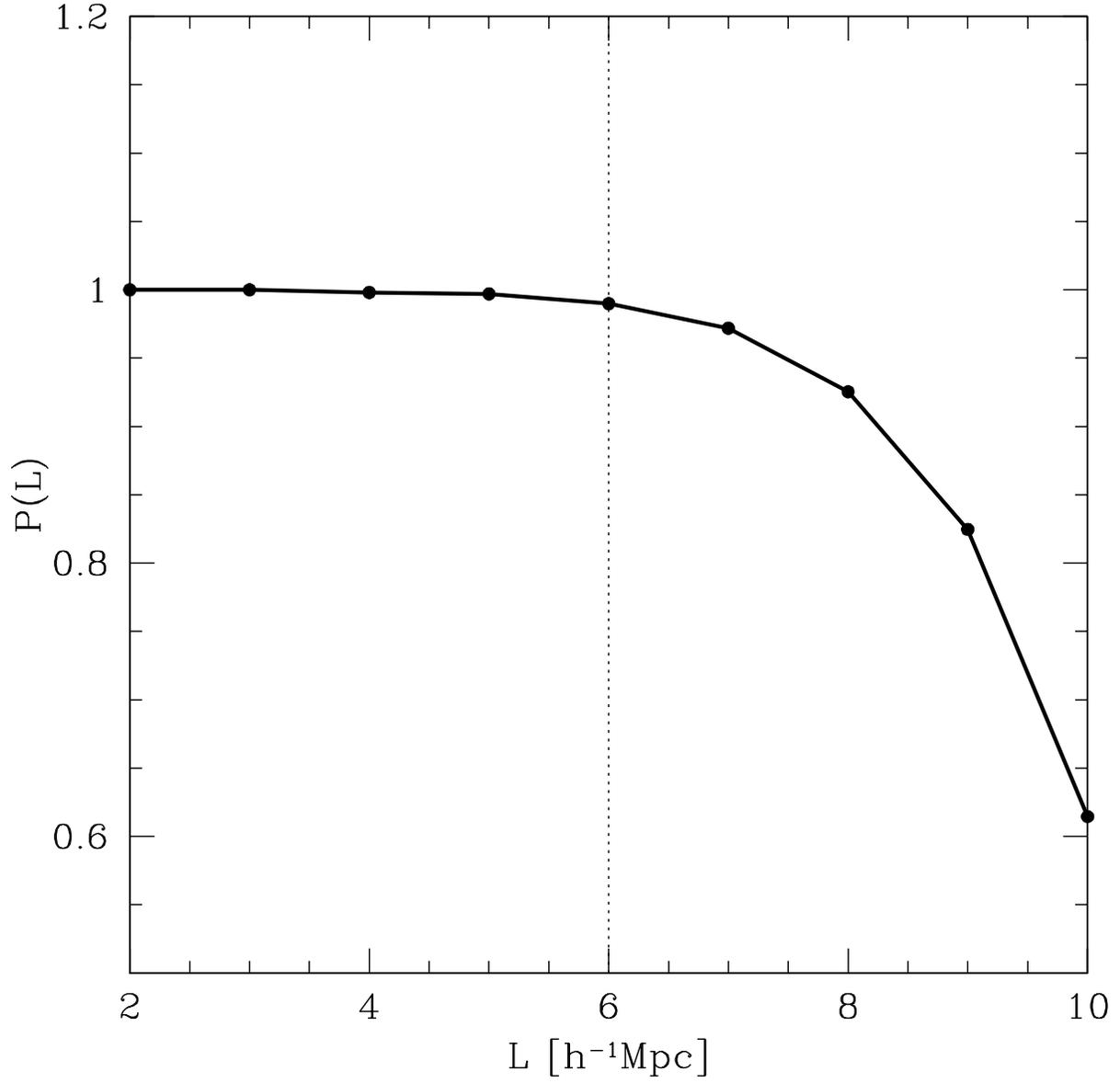}
\caption{The statistical significance of the linking length for the 
FOF algorithm to find superclusters in the Millennium Run halo catalog.}
\label{fig:lin}
 \end{center}
\end{figure}

Figure \ref{fig:lin} plots the statistical significance $P(L)$, which reveals 
that around $L=6 h^{-1}$ Mpc the statistical significance of finding a 
supercluster reaches $99\%$ confidence level.  Therefore, we set the linking 
length for the FOF algorithm at $L=6h^{-1}$Mpc which corresponds to the 
linkage parameter $b=0.35$. 

A total of $4014$ superclusters are found. Among them, a total of $382$ 
superclusters are found to consist of more than $5$ members. 
The mean mass of these $382$ superclusters are found to be 
$\bar{M}_{s}=4.2\times 10^{14}h^{-1}M_{\odot}$.  Using these superclusters, 
we measured the inertia momentum tensor of each supercluster. Rotating 
the system into the principal axis frame, we find the three eigenvectors 
of the inertia momentum tensor of each supercluster. 

\subsection{Alignments between Void Spin and Supercluster Principal Axes}

Now that the spin axes of voids and the principal axes of the superclusters 
are all determined, we calculate the squares of the cosines of the angles 
between the the void spin axes and the three axes of the superclusters as 
a function of the separation distance, $r$. 
Let $\hat{\bf y}^{\alpha},\hat{\bf y}^{\beta},\hat{\bf y}^{\gamma}$ 
be the three unit eigenvectors of the supercluster inertia momentum tensor, 
which represent the supercluster major, intermediate, and minor axes, 
respectively. Basically, we calculate the following three correlation 
functions using the selected void-supercluster pairs:
\begin{eqnarray}
\label{eqn:alpha}
\omega^{\alpha}(r) &\equiv& 
\langle\vert\hat{\bf J}\cdot\hat{\bf y}^{\alpha}\vert^{2}\rangle(r) 
- \frac{1}{3},\\
\label{eqn:beta}
\omega^{\beta}(r) &\equiv& 
\langle\vert\hat{\bf J}\cdot\hat{\bf y}^{\beta}\vert^{2}\rangle(r) 
- \frac{1}{3},\\
\label{eqn:gamma}
\omega^{\gamma}(r) &\equiv& 
\langle\vert\hat{\bf J}\cdot\hat{\bf y}^{\gamma}\vert^{2}\rangle(r)
- \frac{1}{3}.
\end{eqnarray}
If there is no alignment, then these cross-correlation function will be 
just zero.

\begin{figure}
  \begin{center}
   \plotone{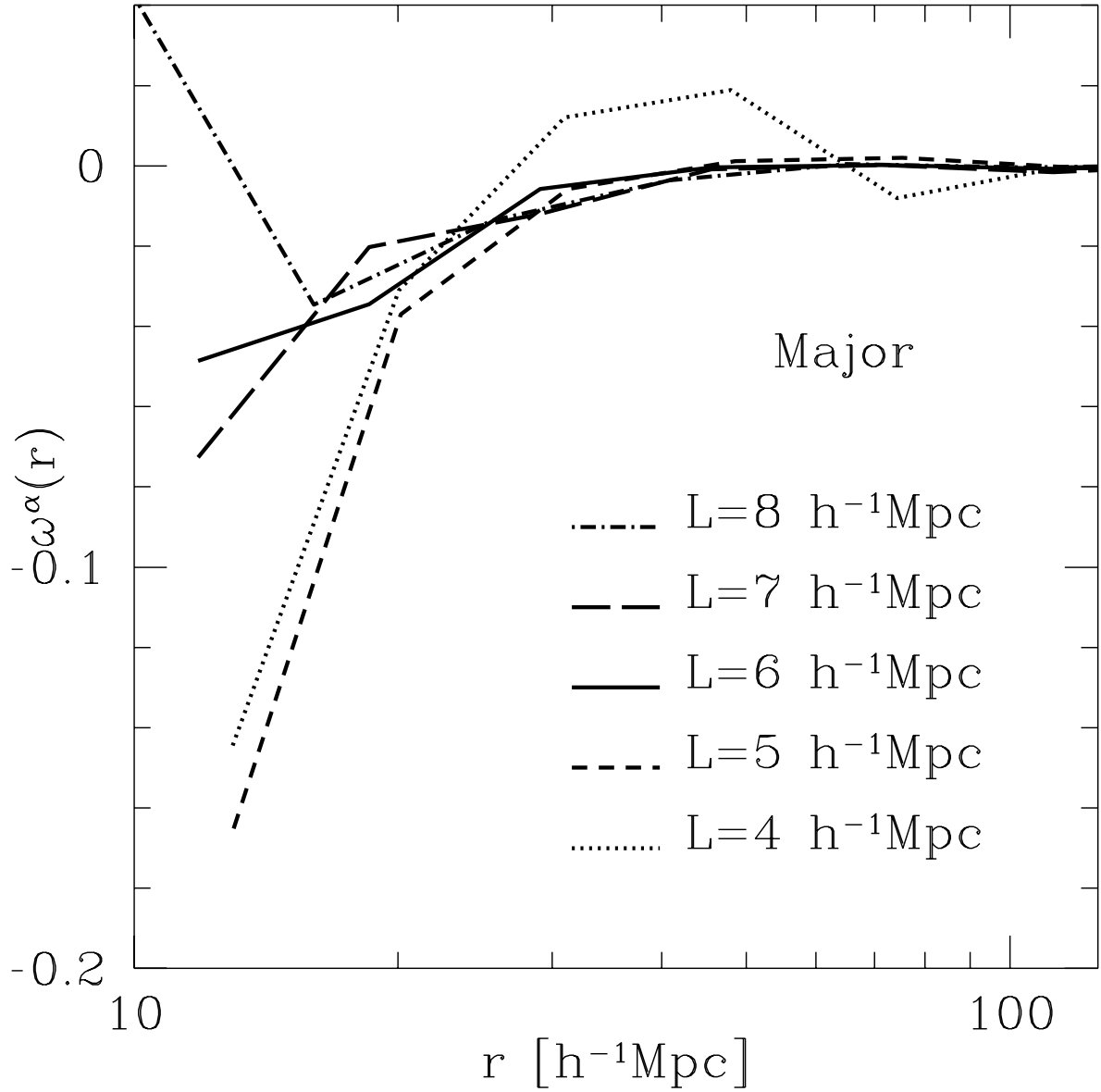}
\caption{Numerical result of the cross-correlation between the void spin 
axes and the major axes of the neighboring superclusters as a function of 
separation distance for the five different cases of the linking length, $L$.}
\label{fig:varma}
 \end{center}
\end{figure}

\begin{figure}
  \begin{center}
   \plotone{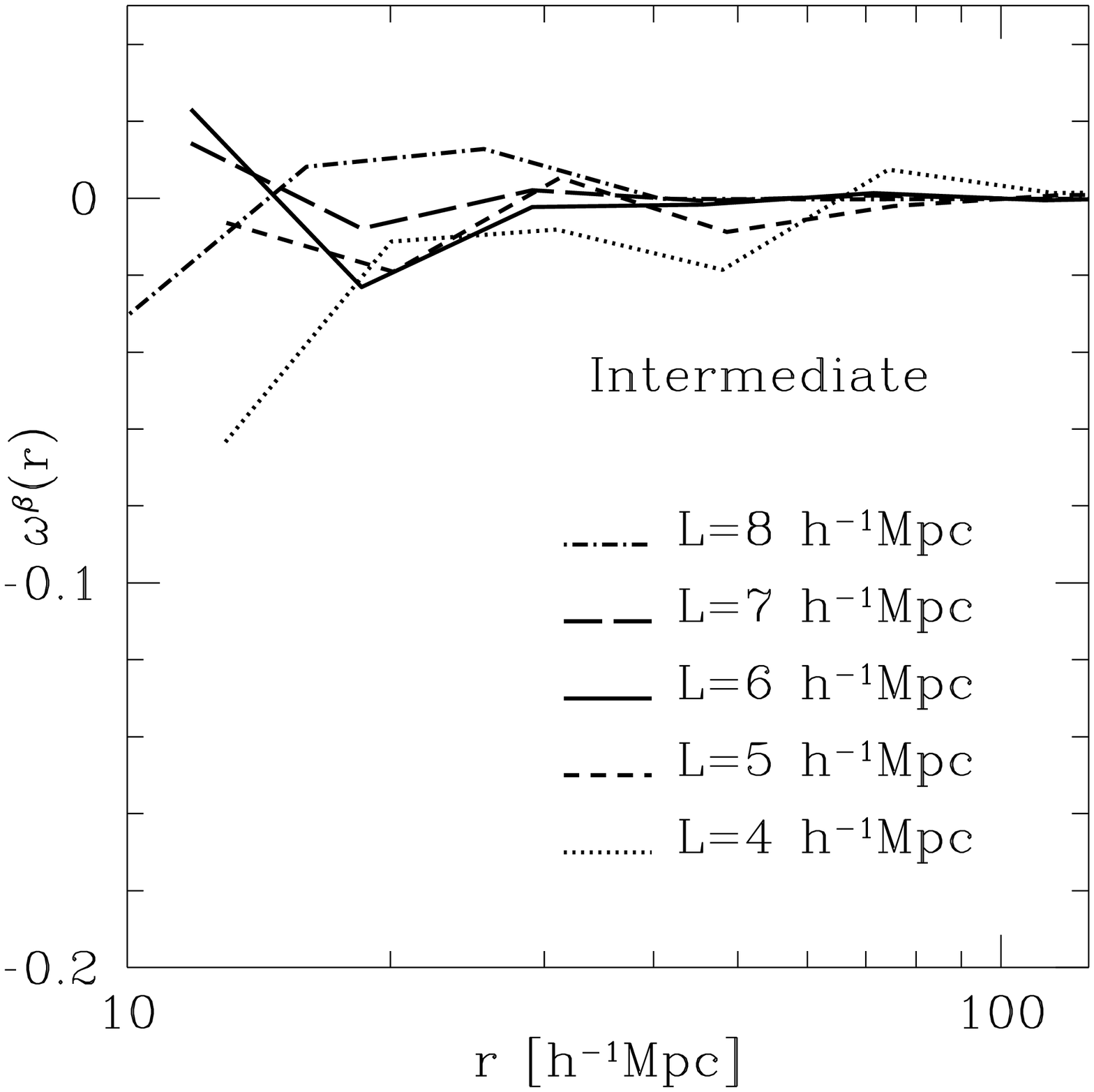}
\caption{Same as Figure \ref{fig:major} but with the supercluster 
intermediate axes.}
\label{fig:varin}
 \end{center}
\end{figure}

\begin{figure}
  \begin{center}
   \plotone{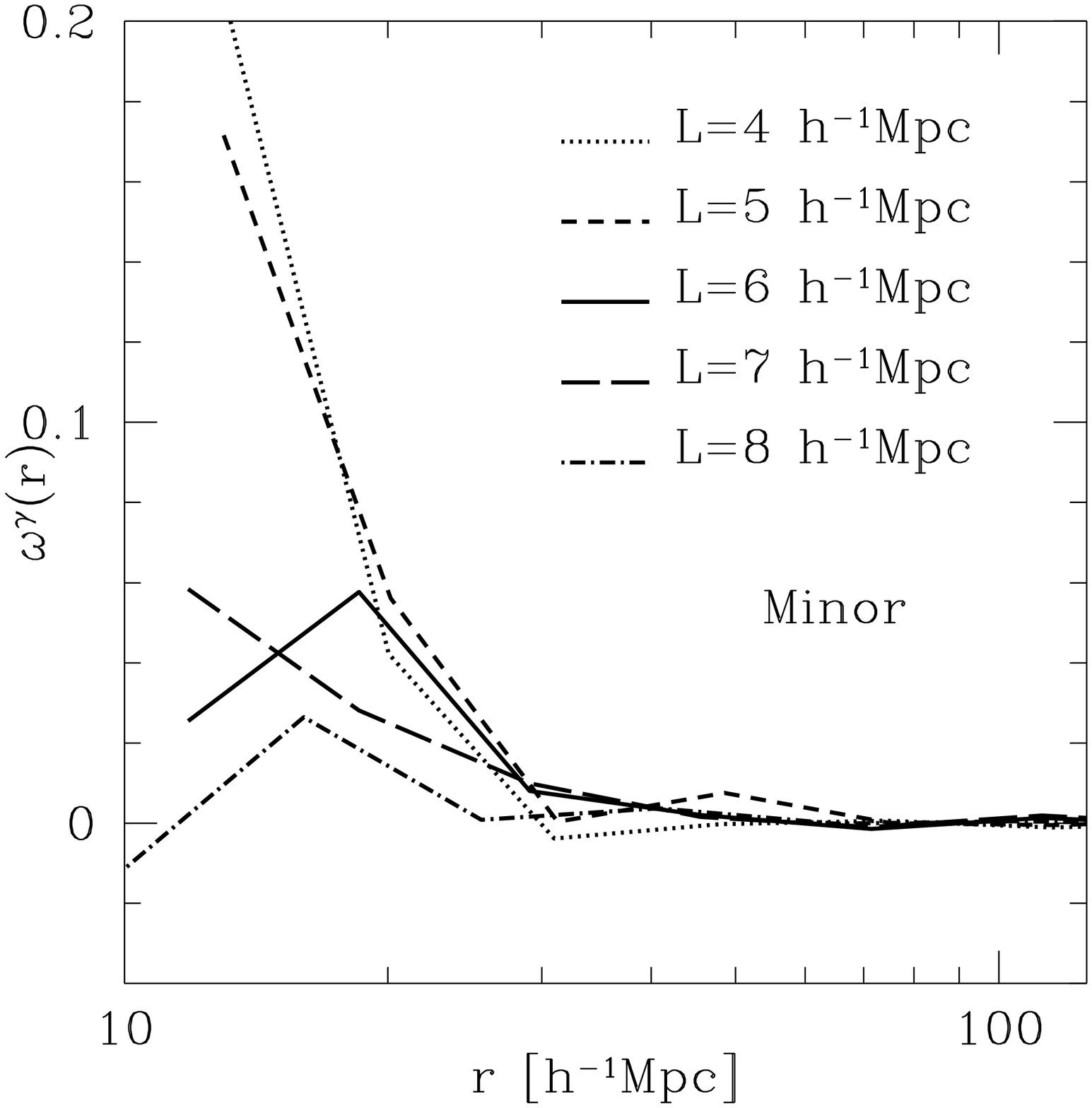}
\caption{Same as Figure \ref{fig:major} but with the supercluster 
minor axes.}
\label{fig:varmi}
 \end{center}
\end{figure}

To examine how these void-supercluster cross correlations depend on the 
linking length, we repeat the whole process using different values of $L$. 
The three cross-correlations functions between the void spin axes and the 
supercluster major ($\omega^{\alpha}$), intermediate ($\omega^{\beta}$), 
and minor axes ($\omega^{\gamma}$) for the fives different cases of the 
linking lengths are plotted in Figs. \ref{fig:varma}, \ref{fig:varin} and 
\ref{fig:varmi}.  As can be seen, for those values of $L$ that corresponds 
to higher than $90\%$ confidence levels (i.e., $L \le 7h^{-1}$Mpc), there are 
clear consistent anti-alignment and alignment signals between the void spin 
axes and the supercluster major and the minor axes, respectively, within a 
distance of $30h^{-1}$Mpc, and there is consistently no alignment signal 
with the supercluster intermediate axes.

We provide  physical explanations to this phenomena in \S 3 within the 
analytic framework proposed originally by \citet{lee-pen01}.

\section{ANALYTIC PREDICTION}

\subsection{Review of Spin-Shear and Direction-Shear Correlations}

According to the Lee-Pen formalism based on the linear tidal torque theory 
\citep{dor70,whi84}, the unit spin vector 
$\hat{\bf J}\equiv (\hat{J}_{i})$ 
of a bound halo is correlated with the unit traceless tidal shear tensor 
$\hat{\bf T}\equiv (\hat{T}_{ij})$ as  
\begin{equation}
\label{eqn:hspincorr}
\langle \hat{J}_{i}\hat{J}_{j}\vert\hat{\bf T}\rangle = 
\frac{1+a}{3}\delta_{ij} - a\hat{T}_{ik}({\bf x})\hat{T}_{kj}({\bf x}),
\end{equation}
where $a$ is the spin-shear correlation parameter in the range of 
$[0,3/5]$. It represents the strength of the correlation between $\hat{\bf J}$ 
and $\hat{\bf T}$: If $a=3/5$, the correlation is strongest. If $a=0$, 
there is no correlation. Here, the unit tidal tensor $\hat{\bf T}$ is 
intrinsic and local, defined at the halo position, ${\bf x}$.

\citet{lee-pen01} also suggested the following formula for the 
correlation between the direction vector to the nearest neighbor 
$\hat{\bf y}\equiv (\hat{y}_{i})$ with the unit traceless tensor 
$\hat{\bf T}$ as 
\begin{equation}
\label{eqn:hdencorr}
\langle \hat{y}_{i}\hat{y}_{j}\vert\hat{\bf T}\rangle = 
\frac{1-b}{3}\delta_{ij} + b\hat{T}_{ik}({\bf x})\hat{T}_{kj}({\bf x}), 
\end{equation}
Here the parameter, $b$ in the range of $[-1,1]$ represents the strength of 
the correlation between $\hat{\bf y}$ and $\hat{\bf T}$. 
Note the difference in the range between the two correlation 
parameters, $a$ and $b$. The maximum value of $b$ is $1$ while that of 
$a$ is $3/5$, less than unity. It reflects the fact that a perfect 
alignment is allowed between $\hat{\bf y}$ and $\hat{\bf T}$  
but not between $\hat{\bf J}$ and $\hat{\bf T}$. 

Anyway, a crucial implication of equations (\ref{eqn:hspincorr}) and 
(\ref{eqn:hdencorr}) is that the spin axes of halos should be closely 
correlated with the directional geometry of the nearby halo distribution 
since $\hat{\bf J}$ and $\hat{\bf y}$ are both correlated with the 
tidal field. In \S 3.2, we extrapolate the validity of 
equations (\ref{eqn:hspincorr}) and (\ref{eqn:hdencorr}) which hold good 
for halos to the voids and superclusters.

\subsection{Modeling Void-Supercluster Alignments}

LP06 have already extrapolated the validity of equation (\ref{eqn:hspincorr}) 
to unbound voids, assuming that the spin-shear correlation parameter $a$ has 
the maximum value of $3/5$ for the case of voids whose spin is defined  as 
(\ref{eqn:voidspin}):
\begin{equation}
\label{eqn:spincorr}
\langle \hat{J}_{i}\hat{J}_{j}\vert\hat{\bf T}\rangle = 
\frac{8}{15}\delta_{ij} - \frac{3}{5}
\hat{T}_{ik}({\bf x}_{v})\hat{T}_{kj}({\bf x}_{v}),
\end{equation}
where $\hat{\bf T}$ is now defined at the void center, ${\bf x}_{v}$. 

Now, we attempt to extrapolate the validity of equation (\ref{eqn:hdencorr}) 
to the alignments between the void spin axes and the supercluster principal 
axes. The superclusters are conspicuously elongated along local filaments 
where the dark matter are preferentially located. The nearest neighbors are 
most likely to be found in the direction along the supercluster major axes. 
But, the filaments are one dimensional structure, collapsed along the 
major and the intermediate principal axes of the local tidal tensors. 
Thus, the direction of local filaments (i.e., the supercluster major axes) 
are in fact anti-aligned with the tidal tensor major axes, and aligned 
with the tidal tensor minor axes.

Using the above logic, we assume the following:
\begin{itemize}
\item
For the direction of the supercluster major axis, $\hat{\bf y}^{\alpha}$, 
the direction-shear correlation parameter $b$ has the minimum 
value of $-1$:
\begin{equation}
\label{eqn:mdencorr}
\langle \hat{y}^{\alpha}_{i}\hat{y}^{\alpha}_{j}\vert\hat{\bf T}\rangle = 
\frac{2}{3}\delta_{ij} - \hat{T}_{ik}({\bf x}_{s})\hat{T}_{kj}({\bf x}_{s}). 
\end{equation}
\item
For the direction of the supercluster intermediate axis,$\hat{\bf y}^{\beta}$, 
the direction-shear correlation parameter $b$ has the value of $0$:
\begin{equation}
\label{eqn:idencorr}
\langle \hat{y}^{\beta}_{i}\hat{y}^{\beta}_{j}\vert\hat{\bf T}\rangle = 
\frac{1}{3}\delta_{ij}, 
\end{equation}
\item
For the direction of the supercluster minor axis, $\hat{\bf y}^{\gamma}$, 
the parameter $b$ has the maximum value of $1$:
\begin{equation}
\label{eqn:rdencorr}
\langle \hat{y}^{\gamma}_{i}\hat{y}^{\gamma}_{j}\vert\hat{\bf T}\rangle = 
\hat{T}_{ik}({\bf x}_{s})\hat{T}_{kj}({\bf x}_{s}), 
\end{equation}
\end{itemize}
Note that in equations (\ref{eqn:mdencorr})-(\ref{eqn:rdencorr}), the unit 
tidal tensor is defined at the supercluster center, ${\bf x}_{s}$.

Using equations (\ref{eqn:spincorr})-(\ref{eqn:rdencorr}), one can derive 
analytically the three correlation functions, $\omega^{\alpha}$, 
$\omega^{\beta}$ and $\omega^{\gamma}$, defined in \S 2.. 
In this derivation, the key part is to calculate the four point shear 
correlation, $\langle\hat{T}({\bf x}_{v})\hat{T}({\bf x}_{v})
\hat{T}({\bf x}_{s})\hat{T}({\bf x}_{s})\rangle$. 
\citet{lee-pen01} calculated this quantity for the case that the two unit 
tidal tensors are defined at the same halo position but smoothed on two 
different scales. What they found is that it is approximated as the density 
auto-correlation \citep[see Appendix I in][]{lee-pen01}.  
In our case, the two unit tidal tensors are not defined at the same 
position but at two different positions, the void and the supercluster 
centers. Therefore, the four-point shear correlation would not be approximated 
as the density autocorrelation. Instead, it may be approximated 
as the density two-point correlation, just like the void spin-spin 
correlation function \citep{lee-par06}.

Hence, using the same approximation used in \citep{lee-par06} but 
considering the fact that the maximum value of the correlation parameter 
$b$ is different from that of $a$, we find that the void-supercluster 
alignments can be approximated as 
\begin{eqnarray}
\label{eqn:aalpha}
\omega^{\alpha}(r) &\approx& 
-\frac{1}{10}\frac{\xi^{2}_{R}(r)}{\xi^{2}_{R}(0)},\\
\label{eqn:abeta}
\omega^{\beta}(r) &\approx& 0,\\
\label{eqn:agamma}
\omega^{\gamma}(r) &\approx& \frac{1}{10}\frac{\xi^{2}_{R}(r)}{\xi^{2}_{R}(0)},
\end{eqnarray}
where $r\equiv\vert{\bf x}_{v}-{\bf x}_{s}\vert$.
Here $\xi_{R}$ is the two point correlation function of the density field 
on the smoothing scale of $R$. For the void-supercluster alignments, the 
smoothing scale should be a minimum Lagrangian radius enclosing a 
void-supercluster pair. Since the separation distance $r$ cannot decrease 
below the sum of the supercluster radius and the void diameter in Lagrangian 
space, the smoothing scale may be written as
\begin{equation}
R = R_{s} + 2R_{v},
\end{equation}
where $R_{s}$ and $R_{v}$ represent the effective radii of a supercluster 
and a void, respectively. 

For the comparison with the numerical results, we relate the Lagrangian 
supercluster and void radii to the observables given in \S 2:
\begin{equation}
\bar{R}_{s} \equiv \left(\frac{3\bar{M}_s}{4\pi\bar{\rho}}\right)^{1/3}, 
\qquad \bar{R}_{v} \equiv (1 + \bar{\delta}_v)^{1/3}\bar{R}_{E},
\end{equation}
where $\bar{\rho}$ is the mean mass density of the universe.

\begin{figure}
  \begin{center}
   \plotone{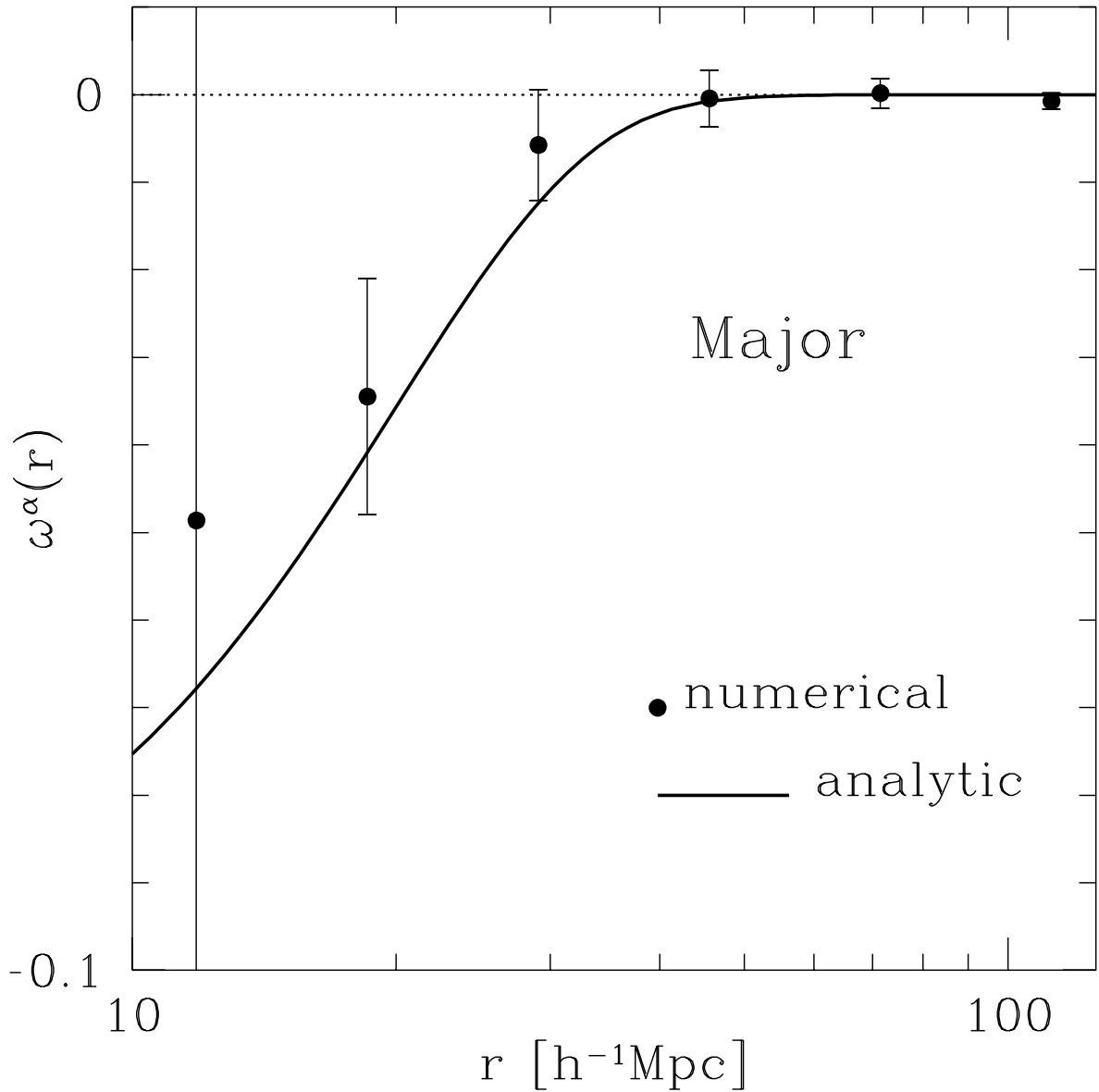}
\caption{Comparison of the analytic cross-correlation between the void spin 
axes and the major axes of the neighboring superclusters (solid line) 
with the numerical result from the Millennium Rum simulation (dots). The 
linking length for the numerical result is set at $6^{-1}$Mpc which 
corresponds to the $99\%$ confidence level for the supercluster identification 
using the FOF algorithm. The errors represent the standard deviation for the 
case of  no alignment. The horizontal line corresponds to the case of no 
alignment}
\label{fig:major}
 \end{center}
\end{figure}

\begin{figure}
  \begin{center}
   \plotone{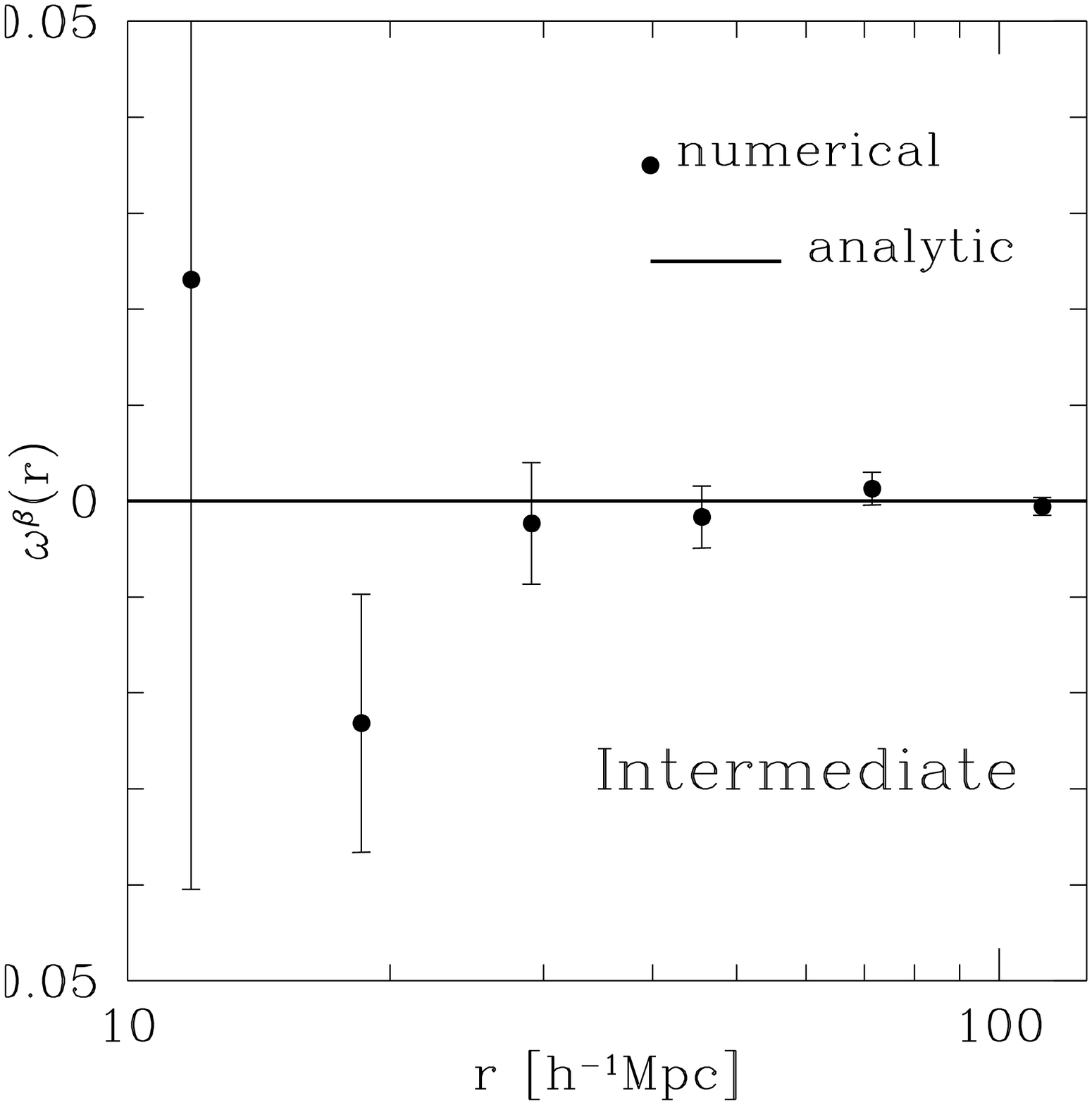}
\caption{Same as Figure \ref{fig:major} but with the supercluster 
intermediate axes.}
\label{fig:inter}
 \end{center}
\end{figure}

\begin{figure}
  \begin{center}
   \plotone{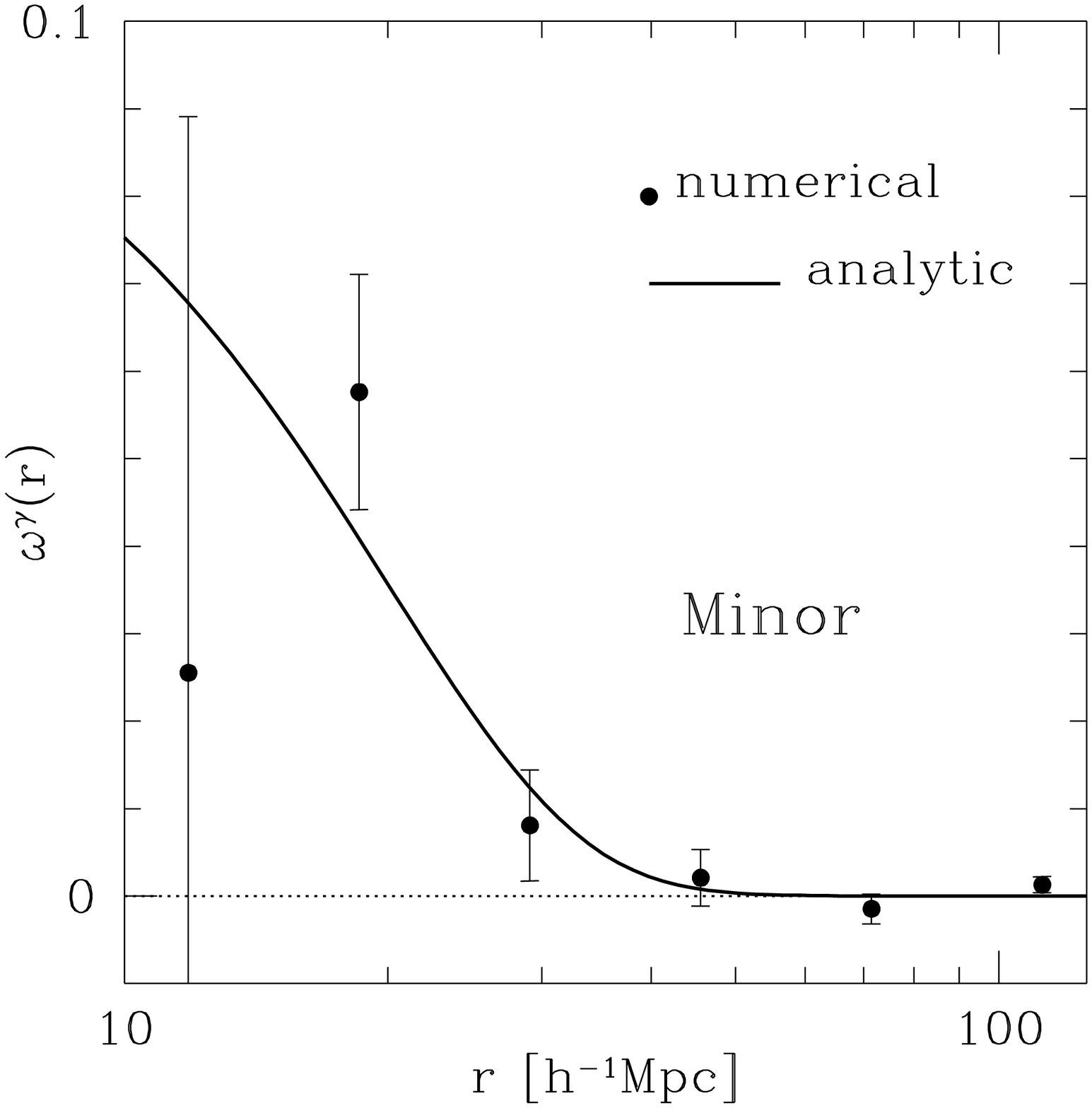}
\caption{Same as Figure \ref{fig:major} but with the supercluster 
minor axes.}
\label{fig:minor}
 \end{center}
\end{figure}

Equations (\ref{eqn:aalpha}), (\ref{eqn:abeta}) and (\ref{eqn:agamma}) are 
plotted in Figs. \ref{fig:major}, \ref{fig:inter} and \ref{fig:minor}, 
respectively, where the numerical results (with $L = 6h^{-1}$Mpc) derived in 
\S 2.3 are also plotted as solid dots with Poissonian errors. For the analytic 
results, we use the transfer function of the initial power spectrum given by 
\citet{bar-etal86} with the cosmological parameters set at the values used in 
the Millennium Run simulations. And, the shape parameter, $\Gamma$ is 
approximated as $\Gamma=\Omega_{m}h$ (V. Springel in private communication). 
As can been obviously seen, the analytic approximations work very well. 
The excellent agreements between the analytical and the numerical results 
imply that the void-supercluster alignments are indeed generated by the tidal 
interactions between the voids and the superclusters.

\section{DISCUSSION AND CONCLUSION}

We have investigated correlations in spatial orientations between voids 
and their neighbor superclusters using the Millennium Run simulation of 
a concordance cosmology.  Adopting the concept of void spin proposed by 
\citet{lee-par06}, we have found for the first time that the void spin 
axes are very strongly correlated with the supercluster minor axes 
within the separation distance of $30h^{-1}$Mpc. Testing how the result 
depend on the choice of void and supercluster definition, we have found 
that our result is quite solid. 

To this numerical phenomena has a physical explanation been provided based 
on tidally generated correlations. Under the assumption that the 
neighboring superclusters are representative of the filamentary 
surrounding matter which wrap and exert tidal forces on the voids, 
we have derived an analytic formula for the alignments between the void 
spin axes and the supercluster minor axes. The analytic prediction has 
turned out to agree with the numerical result very well.

However, it is worth discussing a caveat which the success of our work is 
subject to. This caveat lies in the fact that there is no consensus on how 
to define voids and superclusters, unlike the case of halo-defining.
Using different algorithms could produce different results for correlations 
between voids and superclusters. Although we have shown here that the 
void-supercluster alignments do not strongly depend on the linking length 
of the FOF algorithm and the cut-off mass of the void halos, it will 
definitely necessary to compare between different methods for the void 
and supercluster identifications in the future.

Together with our previous work on the void spin-spin correlation 
\citep{lee-par06}, this new result supports the scenario that the voids 
originate from the initial regions where the tidal effect becomes maximum.
In addition, this new result on the void-supercluster alignments 
demonstrate more directly how the largest scale structure and voids are 
connected in a cosmic web through tidal influences.

A final conclusion is that our work will provide a new clue to describing the 
void-supercluster network in the cosmic web and leads us to have a deeper 
insight into the formation and evolution of the large scale structure of 
the universe.
\acknowledgments

The Millennium Run simulation used in this paper was carried out by the Virgo 
Supercomputing Consortium at the Computing Centre of the Max-Planck Society 
in Garching. We thank V. Springel and G. Lemson for plenty of helps. 
This work is supported by the research grant No. R01-2005-000-10610-0 from 
the Basic Research Program of the Korea Science and Engineering Foundation.

\clearpage
\clearpage
\clearpage
\clearpage
\clearpage
\clearpage

\begin{thebibliography}{100}
\bibitem[Bardeen et al.(1986)]{bar-etal86}
Bardeen, J.M., Bond, J.R.,Kaiser, N., \& Szalay, A.S. 1986, \apj, 304, 15
\bibitem[Basilakos et al.(2006)]{bas-etal06}
Basilakos, S., Plionis, M., Yepes, G., Gottlober, S., Turchaninov, V. 2005, 
\mnras, 365, 539
\bibitem[Colberg et al.(2005)]{col-etal05}
Colberg J. M., Sheth R. K., Diaferio A., Gao L., \& Yoshida N. 2005,
\mnras, 360, 216
\bibitem[Doroshkevich(1970)]{dor70}
Doroshkevich, A. G. 1970, astrofizika, 6, 581
\bibitem[El-Ad \& Piran(1997)]{ela-pir97}
El-Ad, H., \& Piran, T. 1997, \apj, 491, 421
\bibitem[Gottl\"{o}ber et al.(2003)]{got-etal03}
Gottl\"{o}ber, S., Lokas, E. L., Klypin, A., \& Hoffman, Y. 2003,
\mnras, 344, 715
\bibitem[Hoyle \& Vogeley(2002)]{hoy-vog02}
Hoyle, F., \& Vogeley, M. S. 2002, \apj, 566, 641
\bibitem[Hoyle \& Vogeley(2004)]{hoy-vog04}
Hoyle, F., \& Vogeley, M. S. 2004, \apj, 607, 751
\bibitem[Icke(1984)]{ick84}
Icke V. 1984, \mnras, 206, 
\bibitem[Lee \& Park(2006)]{lee-par06}
Lee, J. \& Park, D. 2006, \apj, 652, 1
\bibitem[Park \& Lee(2007)]{lee-par07}
Park, D. \& Lee, J. 2007, submitted to Phys. Rev. Lett.
\bibitem[Lee \& Pen(2001)]{lee-pen01}
Lee, J. \& Pen, U. L. 2001, \apj, 555, 106 
\bibitem[Sahni et al.(1994)]{sah-etal94}
Sahni, V., Sathyaprakah, B. S., \& Shandarin, S. F. 1994, \apj, 431, 20
\bibitem[Shandarin et al.(2004)]{sha-etal04}
Shandarin, S. F., Sheth, J., \& Sahni, V. 2004, \mnras, 353, 517
\bibitem[Shandarin et al.(2006)]{sha-etal06}
Shandarin, S., Feldman, H. A., Heitmann, K., \& Habib, S. 2006,
\mnras, 367, 1629
\bibitem[Sheth \& van de Weygaert(2004)]{she-van04}
Sheth, R. K., \& van de Weygaert, R. 2004, \mnras, 350, 517
\bibitem[Springel et al.(2005)]{spr-etal05}
Springel, V. et al. 2005, \nat , 435, 629
\bibitem[Wray et al.(2006)]{wra-etal06}
Wray, J. J., Bahcall, N. A., Bode, P., Boettiger, C., Hopkins, P. F. 2006, 
\apj, 652, 907
\bibitem[White(1984)]{whi84}
White, S. D. M. 1984, \apj, 286, 38
\end{thebibliography}
\end{document}